\documentclass{ws-ijmpcs}

\begin{document}

\markboth{P.~Bordas, V.~Bosch-Ramon and M.~Perucho}
{Large-scale emission in FR I jets}

%
\catchline{}{}{}{}{}
%

\title{\textbf{LARGE-SCALE EMISSION IN FRI JETS}}

\author{POL BORDAS}


\address{Institut f\"ur Astronomie und Astrophysik T\"ubingen, Sand 1, 72076 T\"ubingen, Germany\\
{\textit{INTEGRAL}} Science Data Centre, Universit\'e de Gen\`eve, Chemin d'Ecogia 16, CH--1290 Versoix, Switzerland\\
E-mail: pol.bordas@uni-tuebingen.de\\}

\author{V. BOSCH-RAMON}
\address{Dublin Institute for Advanced Studies, 31 Fitzwilliam Place, Dublin 2, Ireland.\\
E-mail: valenti@cp.dias.ie}

\author{M. PERUCHO}
\address{Departament d'Astronomia i Astrof\'{\i}sica. Universitat de Val\`encia. C/ Dr. Moliner 50, 46100 Burjassot (Val\`encia), Spain\\
E-mail:perucho@uv.es}

\maketitle

\begin{history}
\received{Day Month Year}
\revised{Day Month Year}
\end{history}

\begin{abstract}

The termination structures of the jets of Fanaroff \& Riley (FR) galaxies are observed to produce extended non-thermal emission in a wide frequency range. The study of these structures can provide valuable insights on the conditions for particle acceleration and radiation at the shock fronts. We have studied the thermal and non-thermal emission that can be expected from the jet termination regions of Fanaroff \& Riley type I sources. The broadband emssion from these galaxies has been recently extended to include the high-energy gamma-ray domain, owing to the \textit{Fermi} detection of Cen~A lobes. Exploring the physics behind the jet/medium interactions in FRI can provide valuable insights on the conditions for particle acceleration and radiation in the jet termination shocks. Making use of the results of a fully relativistic numerical simulation code of the evolution of a FRI jet we model the expected radiative output and predict spectra and lightcurves of both thermal and non-thermal emission at different source ages.

\end{abstract}

\ccode{PACS numbers: 98.54.Gr, 98.58.Fd,  95.30.Gv}

\section{Introduction}

Fanaroff-Riley sources of type I and II (FRI-II) are radio loud Active Galactic Nuclei (AGN) displaying large-scale jets that interact with the surrounding ISM/ICM medium.\cite{fr74}. The study of the extended emission in FRI/II sources can be used to derive the properties of the flows and the surrounding medium, and to determine the conditions for particle acceleration. In this regard a simple evolutionary model can be used to conclude that thermal X-ray emission produced in the shell could be detected only for dense enough media even for powerful FRII jets,\cite{hrb98} whilst thermal MeV radiation from the cocoon region would be significant only for sources with ages $\ll 10^7\,\rm{yrs}$.\cite{kki07}. Non-thermal radiation can also be produced in the jet/medium interaction regions. X-ray synchrotron and/or IC emission has been observed in several sources (see e.g. \refcite{ka03,cr09}), and the {\it Fermi} Collaboration recently reported the detection of extended GeV emission in the radio lobes of Cen~A, demonstrating that acceleration up to very high energies is indeed taking place in the disrupted jet region.\cite{ab10a}  Numerical simulations, on the other hand, have demonstrated to be a powerful tool to study the jet/medium interactions. Perucho \& Mart\'i (2007) performed a simulation aimed to test the FRI jet evolution paradigm.\cite{pm07,b84} We computed the thermal and non-thermal emission from FRI sources obtained with a radiative model adapted from Bordas et al. (2009) that takes as input the results of those simulations as input and the radiation model used to characterize the conditions of the shocked regions.\cite{br09} Below, we briefly introduce the numerical simulations and the radiation model used. We then discuss the obtained results and the relevance of the thermal and non-thermal radiation at different energy bands. For a more extended discussion, the reader is referred to Perucho \& Mart\'i (2007) and Bordas et al. (2011).\cite{pm07, br11}

\section{Hydrodynamical simulations and radiation model}

The results of the numerical simulations of Perucho \& Mart\'i (2007) have been used to characterize the physical properties of the jet/medium shocked regions.\cite{pm07} The jet is injected in the numerical grid at $500\,\rm{pc}$ from the active nucleus, with a radius of $60\,\rm{pc}$. The ambient  medium, composed by neutral hydrogen, has a profile in pressure (see, e.g., \refcite{hr02}). The jet, leptonic in composition, has an initial velocity $v_{\rm j0}=0.87\,c$, a density ratio with respect to the ambient $\rho_{\rm j0}/\rho_{\rm a0}=10^{-5}$, a pressure ratio with the ambient $P_{\rm j0}/P_{\rm a0}\simeq 8$, and temperature $4\times10^9\,\rm{K}$, resulting in a kinetic luminosity $L_{\rm j}=10^{44}$~erg~s$^{-1}$. Further details on the simulation code can be found in \refcite{pm07}. Regarding the radiative code, we used a simplified one-zone model to study the emission properties of both shell and cocoon regions. The injected non-thermal luminosity is 10\% of the total jet kinetic luminosity $L_{\rm j}$. The magnetic field $B$ has been fixed taking the magnetic energy density to be 10\% of the ram/thermal pressure. Concerning particle acceleration, the recollimation shock is assumed to be the accelerator of particles in the cocoon region. We use a relativistic approach in this case, with an acceleration rate  $\dot{E}=\eta\,qBc$ with $\eta=0.1$. For the bow shock, we have adopted instead a non-relativistic approach in view of the lower shock velocities $\sim (1-2)\times 10^8$~cm~s$^{-1}$. The particle energy distribution at a given time, $N(E,t_{\rm src})$, is calculated considering the injection ($\propto E^{-p}$, with $p=2.1$), at any time $t_{\rm src}$. Maximum particle energies, $E_{\rm max}(t)$, vary also along time, through the magnetic field, the accelerator size and the shock velocity dependence. Further details on the radiative code can be found in \refcite{br09} and \refcite{br11}.

\section{Results}
\subsection{Non-thermal emission}

\begin{center}
\begin{figure*}
\includegraphics[width=1.0\textwidth]{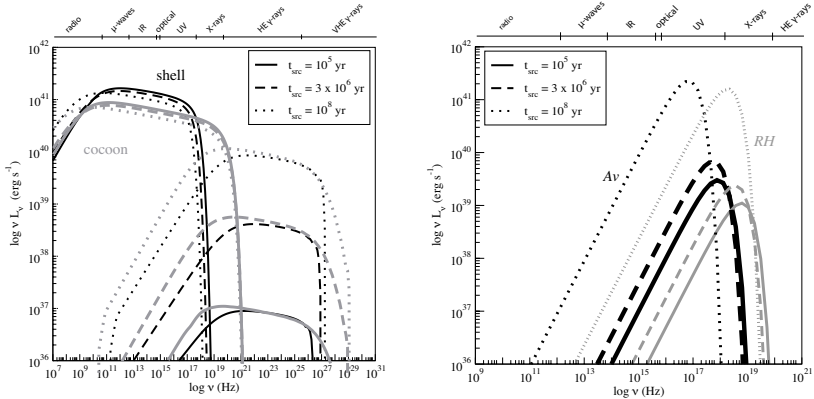}
\caption{ \small{SEDs of the synchrotron and IC (\textit{left}) and Bremsstrahlung emission (\textit{right}) for $ t_{\rm src} = 10^5$ (solid line), $3\times 10^6$ (long-dashed line) and $10^8$~yr (dotted line). Both shell and cocoon contribute to the overall non-thermal emission. Thermal radiation is computed only in the shell, and has been splitted into one component corresponding to the average shell properties ($Av$, black thick), and another one produced in the shell region with conditions similar to those of Rankine-Hugoniot ($RH$, green thin). Only one jet/medium interaction region is shown.}}
\label{thermal_and_non_thermal_SED}
\end{figure*}
\end{center}

\begin{figure*}
\includegraphics[width=0.9\textwidth]{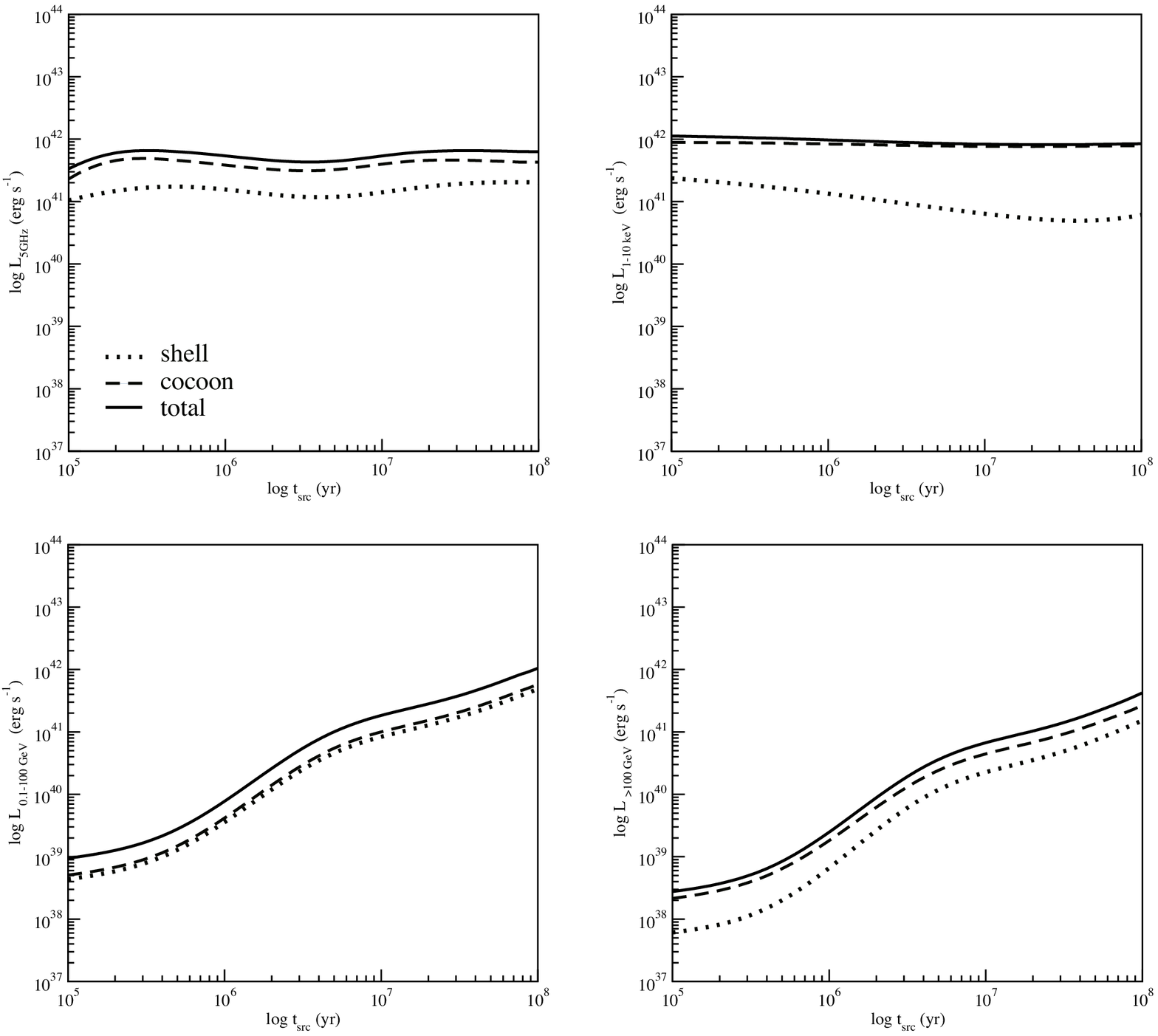}
\caption{ \small {Computed non-thermal lightcurves of the radio (${\rm 5~GHz}\times L_{\rm 5~GHz}$, top-left), 
X-ray (bolometric: 1--10~keV, top-right) and gamma-ray emission 
(bolometric: 0.1--100~GeV, bottom-left; bolometric: $>100$~GeV: bottom-right) in the age range $t_{\rm src}=10^5-10^8$~yr.}}
\label{nonthermal_lc}
\end{figure*}

\noindent The non-thermal spectral energy distributions (SED) for the cocoon and the shell, at $t_{\rm src}=10^{5}$, $3 \times 10^{6}$ and $10^{8}$~yr, are shown in the left and center panels of Fig.~\ref{thermal_and_non_thermal_SED}. The obtained radio and X-ray synchrotron luminosities in both regions are at the level of $2\times 10^{41}$~erg~s$^{-1}$. The approximate constancy of the X-ray luminosities with time is due to the fact that particles have reached the steady state at $t_{\rm src}$ through synchrotron cooling and the assumed constancy of $L_{\rm nt}$. The decrease of $B$ with time, and therefore the growth of $t_{\rm syn}$, is compensated by the increase of time available for cooling. The IC luminosity grows as long as this process becomes more efficient compared to synchrotron and adiabatic cooling, which is shown by the decrease of $u_{B}/u_{\rm rad}$ from $\approx 5 \times 10^{3}$ ($10^5$~yr) to 4 ($10^{8}$~yr). The cocoon and the shell have similar HE luminosities, but the cocoon is a few times brighter at VHE than the shell due to the higher maximum energies that particles can attain in the former. In both regions the bolometric IC luminosities grow similarly with time, reaching $\sim 10^{42}$ and $10^{41}$~erg~s$^{-1}$  at HE and VHE, respectively. The lightcurves for the luminosities in radio (at 5~GHz), X-rays (1--10~keV), HE (0.1-100~GeV) and VHE (0.1--100~TeV), for the cocoon and the shell, are presented in Fig.~\ref{nonthermal_lc}. They show in more detail the time behavior of the non-thermal radiation at different wavelengths discussed above. The complex and smooth shape of the lightcurves, most clear for the HE and the VHE emission, is a consequence of the hydrodynamical evolution of the whole interaction structure propagating in an inhomogeneous external medium.

\subsection{Thermal emission}

\noindent Thermal emission has been studied for the shell, much denser than the cocoon. We have only considered thermal Bremsstrahlung emission, although we
note that recombination radiation could also be significant (with luminosities $L_{\rm rec} \gtrsim 10^{-21} \, n_{\rm e} n_{\rm ion} \, T^{-1/2} \, V_{\rm sh}
 \, Z^{4}$~erg~s$^{-1}$, where $V_{\rm sh}$ is the shell volume filled with electrons and ions with particle densities $n_{\rm e}$ and
$n_{\rm ion}$, respectively, and $eZ$ is the ion charge \cite{cooper1966}). We considered two separate regions: one, cooler (ultraviolet -UV-/soft X-rays) but brighter, corresponds to
the averaged shell conditions ($Av$), and another one, fainter but hotter (hard X-rays), corresponds to a region close to the apex of the bow shock, with
properties close to those given by the Rankine-Hugoniot ($RH$) conditions. The volume of the latter region is about $3-4$\% of that of the whole shell.
Figure~\ref{thermal_and_non_thermal_SED}, right panel, shows the thermal SEDs for $t_{\rm src}=10^{5}$, $3 \times 10^{6}$ and $10^{8}$~yr. The slowdown of the
bow shock and the velocity dependence of the postshock temperature, $\propto v_{\rm bs}^2$, leads to a decrease in the peak of the thermal emission with time.
However, the overall luminosity increases as the source gets older, from $10^{39}$~erg~s$^{-1}$ to few times $10^{41}$~erg~s$^{-1}$, since the Bremsstrahlung
time-scale is $t_{\rm Bremss} = 10^{3} T^{1/2} n_{\rm e}^{-1} >> t_{\rm src}$ at any source age, with the shell and the hot postshock region components
peaking from soft X-rays to UV and from hard to soft X-rays, respectively. Thermal Bremsstrahlung increases from $t_{\rm src}=10^{5}$ to $\sim 10^{6}$~yr,
then the luminosity slightly decreases until $t_{\rm src}\sim 3\times 10^{6}$~yr, when a transition in the external medium from the denser galaxy core to the
rarefied galaxy group medium is produced, and increases again afterwards. The component $Av$ dominates the thermal bolometric luminosity in young sources, but
the component $RH$ becomes similarly bright at $t_{\rm src}\sim 10^{8}$~yr.

\section{Discussion}

Our results can be discussed on the ground of the thermal and non-thermal fluxes predicted at different energy bands. In radio, the cocoon dominates with fluxes up to $\sim few \times 10^{-12}\,(d/100~$Mpc$)^{-2}$~erg~cm$^{-2}$~s$^{-1}\sim 10$~Jy at 5~GHz from a region of few times $10'\,(d/100~$Mpc$)^{-1}$ of angular size. This radio fluxes are similar to those observed for instance in 3C~31 and 3C~15,\cite{a92,ka03}.The shell radio flux is sligthly below the cocoon one, although limb brightening may enhance its detectability. The radio lightcurve is quite steady, since the accumulation of electrons compensates the weakening of the magnetic field with $t_{\rm src}$.

At X-rays, the different thermal emitting regions in the shell would lead to a bolometric flux from $\sim 10^{-15}\,(d/100~$Mpc$)^{-2}$ ($10^5$~yr) to a few times $10^{-13}\,(d/100~$Mpc$)^{-2}$~erg~cm$^{-2}$~s$^{-1}$ ($10^8$~yr) in 3C 15. The hard X-rays would be located at the apex of the bow shock, with a typical angular size of a few $1'\,(d/100~$Mpc$)^{-1}$ (for $t_{\rm src}\sim 10^8$~yr), whereas lower energy X-rays would come from the whole shell, with of a few $10'\,(d/100~$Mpc$)^{-1}$. Regarding non-thermal X-rays, the cocoon dominates the total output, with fluxes $\sim 10^{-13}\,(d/100~$Mpc$)^{-2}$~erg~cm$^{-2}$~s$^{-1}$, although limb brightening effects may again increase the shell brightness. In fact, in the case of Cen~A, the shell seems to be the dominant source of non-thermal X-rays.\cite{cr09} This difference could be explained by a higher $E_{\rm max}$ in the shell of that source. The non-thermal X-ray fluxes $\sim 10^{-14}$~erg~cm$^{-2}$~s$^{-1}$ of 3C 15 (see e.g. \refcite{ka03}) imply a non-thermal luminosity of $\approx 1.3 \times 10^{41}$~erg~s$^{-1}$ at 300~Mpc, in good agreement with the values predicted here. Furthermore, synchrotron emission concentrated around the recollimation shock is compatible with the large-scale jet X-ray emission found in 3C~31.\cite{hr02} If a strong recollimation shock is indeed the origin of these large-scale jet X-rays, then the hypothesis that jet disruption in 3C~31 is caused by shock triggered instabilities is favored against stellar wind mass-load.\cite{lb02} Finally, we note that for sources older than those considered here and/or lower $B$-values, the synchrotron emission would be less relevant and IC would dominate the X-ray output.


The moderate velocities of the bow shock found in the numerical simulations prevents thermal photons from reaching energies as high as those discussed in \refcite{kki07}. In the gamma-ray domain, the SED is close to flat at HE, and becomes steeper at VHE. We have not accounted for the gamma-ray absorption induced by the extragalactic background light, which would become significant at distances larger than 100~Mpc. The overall emission increases with time mainly due to the increasing efficiency of the IC channel (considering the cosmic microwave background only) as compared with synchrotron and adiabatic losses. Gamma-ray fluxes for a source with $t_{\rm src}\sim 10^8$~yr are around $\sim 10^{-12}\,(d/100\,{\rm Mpc})$~erg~cm$^{-2}$~s$^{-1}$. At HE, such a source may require very long exposures to be detected by, e.g., {\it Fermi}, unless it is very nearby (as it is the case of Cen~A, see \refcite{ab10a}) or shows a higher non-thermal efficiencies and/or jet power. At VHE, the fluxes could be observed by the current Cherenkov telescopes, although the extension of the source, of tens of arcminutes at 100~Mpc, and the steep spectrum above $\sim 100$~GeV, may make a detection possible only after a long exposure. The forthcoming CTA observatory may however allow the detection of VHE emission from FRI jet lobes, and possibly carry out energy-dependent morphological studies.

From our study we conclude that, for moderate non-thermal luminosities, radio lobes of FRI radio galaxies are good candidates to be detected in the whole spectral range, with the radiation appearing extended in most of the energy bands. Soft X-rays will be likely dominated by synchrotron emission up to ages $\sim 10^8$~yr, with IC tending to be dominant for older sources. Thermal X-rays seem unavoidable and may dominate in hard X-rays in old sources even if a non-thermal component is present. The low surface brightness may require long observation times for the detection in X- and gamma-rays, although the steady nature of these sources can help in this regard.

\end{document}